# Phase Evolution and Substrate-Dependent Nucleation of Quartz GeO$_2$ Films Grown by MOCVD on r- and c-Plane Sapphires

Botong Li [1, †, *], Imteaz Rahaman[1, †], Hunter Ellis[1], Bobby G. Duersch[2], Kathy Anderson[3], Kai Fu[1, *]

[1]*Department of Electrical and Computer Engineering, The University of Utah, Salt Lake City, UT 84112, USA*

[2]*Electron Microscopy and Surface Analysis Laboratory, The University of Utah, Salt Lake City, UT 84112, USA*

[3]*Utah Nanofab, Price College of Engineering, The University of Utah, Salt Lake City, UT 84112, USA*

[*]Corresponding author: Botong Li (E-mail: u1540316@utah.edu); Kai Fu (E-mail: kai.fu@utah.edu)

[†]Authors contributed equally

**Abstract:** Ultrawide-bandgap (UWBG) semiconductors, such as GeO$_2$, are gaining significant attention for their potential in high-performance applications, particularly in piezoelectric devices. Despite extensive research, a comprehensive understanding of the growth dynamics and phase evolution of GeO$_2$ films via metal-organic chemical vapor deposition (MOCVD) remains insufficient. In this study, we investigate the growth behavior and morphological evolution of GeO$_2$ thin films on r-plane and c-plane sapphire substrates for the MOCVD growth process. The temporal evolution of crystallization and the amorphous-to-quartz phase transition are systematically elucidated for the first time. As growth time increases, the spherulitic quartz patterns expand in size, and elevated growth temperatures are found to enhance the crystallization rate. Distinct morphological symmetries emerge depending on the substrate orientation: quadrangular patterns on r-plane sapphire and hexagonal patterns on c-plane sapphire. Atomic force microscopy reveals that these spherulitic domains exhibit pyramid-like surface topography, consistent with volumetric contraction during the amorphous-to-quartz phase transition. These findings offer new insights into the phase evolution and substrate-dependent crystallization behavior of GeO$_2$ films grown by MOCVD.

**Keywords:** GeO$_2$; MOCVD; Phase evaluation; c-Al$_2$O$_3$; r-Al$_2$O$_3$

**Introduction:** UWBG semiconductors such as β-Ga$_2$O$_3$, AlGaN, and diamond are highly valued for their exceptional breakdown strength and power-handling capabilities, making them ideal for high-performance applications like power transistors, ultraviolet photodetectors, and gas sensors [1–5]. However, each of these materials faces inherent limitations: β-$Ga_2O_3$ struggles with poor thermal conductivity and a lack of p-type doping[6–8], AlGaN experiences reduced hole mobility and difficulties with acceptor activation at high Al content [9,10], and diamond encounters challenges in wafer-scale production and doping [11,12]. These limitations have led researchers to explore alternative UWBG oxides, with $GeO_2$ emerging as a particularly promising candidate, specifically due to its ambipolar dopability. $GeO_2$ shows immense potential in semiconductor electronics, energy storage, and optical technologies, where it could contribute to the design and development of new device architectures.

Despite being studied since the 1950s [13], the fascinating properties of $GeO_2$ and polymorphs have continued to drive research. However, synthesizing its various polymorphs, which have relatively close formation energies, remains a challenge, making it difficult to fully control the growth dynamics and phase formation, especially for epitaxy. This complexity is particularly evident when considering the two primary phases of $GeO_2$: the rutile/cristobalite (tetragonal) phase [14] and the quartz (hexagonal) phase. According to the phase diagram of $GeO_2$, the rutile phase (r-$GeO_2$) is the most stable, yet the metastable hexagonal phase (q-$GeO_2$) is more commonly observed due to its lower density ($\rho \approx 4.3$ g/cm$^3$) [16] compared to the rutile phase ($\rho \approx 6.27$ g/cm$^3$)[17], requiring less energy to form [18]. Similar to $SiO_2$, the quartz phase of $GeO_2$, with its trigonal symmetry (space group $P3_12_1$ or $P3_22_1$), presents an alternative with significant piezoelectric potential. While α-quartz $SiO_2$ is widely used in piezoelectric applications, the α-quartz phase of $GeO_2$ offers superior properties, such as a larger lattice size ($a$ = 4.989 Å and $c$ = 5.653 Å) compared to lattice of $SiO_2$ ($a$ = 4.916 Å and $c$ = 5.408 Å) [19], along with higher piezoelectric coefficients of $d_{11}$ = 6.2 ± 0.3 pC/N and $d_{14}$ = 2.7 ± 0.5 pC/N [20], compared to $d_{11}$ = 2.31 pC/N and $d_{14}$ = 0.73 pC/N for $SiO_2$ [21] and better thermal stability.

Significant progress has been made in the thin film growth of quartz $GeO_2$. Zhou *et al.* successfully grew $GeO_2$ films on $Al_2O_3$ (0001) substrates by pulsed-laser deposition (PLD), annealing at 830°C, 880°C, and 930°C under 200 mbar oxygen pressure for 30 minutes [22]. Their work demonstrated the crystallization of pure $GeO_2$ into the quartz structure, with the best quality on sapphire substrates. The study highlighted dendritic and single-crystal growth with lattice rotation, as well as the influence of different substrates, like $MgAl_2O_4$ and STO, on nucleation mechanisms and crystallization patterns, with TiO$_2$ buffer layers improving growth rates and preventing unwanted reactions. Nalam *et al.* used radio-frequency magnetron sputtering to deposit GeO$_2$ films on silicon (100) and quartz substrates [23]. Their work successfully demonstrated the fabrication of nanocrystalline GeO$_2$ in pure α-quartz and rutile phases, with controlled synthesis parameters, revealing phase-pure tetragonal and hexagonal forms and enhanced optical properties, suggesting its potential as a UWBG semiconductor. Nam *et al.* investigated the effect of post-depositional heat treatments on sputtered quartz-GeO$_2$ thin films across 300–900°C and various atmospheres, analyzing how the microstructure and composition influence optical properties such as transmittance and bandgap [16]. In our previous study, we successfully grew polycrystalline quartz GeO$_2$ on sapphire substrates via MOCVD, systematically adjusting parameters to produce crystalline films on both c-plane and r-plane sapphire [15]. However, the nucleation behavior of GeO$_2$ during MOCVD remains poorly understood, posing a critical barrier to elucidating the growth dynamics and achieving precise phase control. This gap in understanding is particularly significant for GeO$_2$ and other complex oxides that exhibit similar polymorphic competition and kinetic constraints during vapor-phase deposition.

In this study, we systematically investigate the heteroepitaxial growth and nucleation behavior of GeO$_2$ on r-plane and c-plane sapphire substrates. A series of experiments were conducted under different growth temperatures and durations using MOCVD. The surface morphology, crystalline structure, and film evolution were characterized by scanning electron microscopy (SEM), X-ray diffraction (XRD), and atomic force microscopy (AFM) to provide a comprehensive understanding of the formation and growth dynamics of the quartz phase of GeO$_2$ crystal patterns. This work provides new insights into the crystallization pathways of GeO$_2$ and establishes fundamental guidelines for growing quartz GeO$_2$ films via MOCVD.

**Experimental details**

Undoped GeO$_2$ films were deposited using an Agilis metal-organic chemical vapor deposition (MOCVD) system. R-plane and C-plane Al$_2$O$_3$ were used as substrates. Prior to deposition, all substrates underwent a standard cleaning procedure involving sequential ultrasonic baths in acetone, isopropanol, and deionized water. This was followed by a piranha solution treatment (H$_2$SO$_4$: H$_2$O$_2$=3:1) to remove organic residues. Two growth temperatures, 925°C and 880°C, were employed to investigate temperature-dependent growth behavior, while other process parameters were kept constant. The chamber pressure and substrate rotation speed were maintained at 300 Torr and 300 RPM, respectively. Tetraethyl Germane (TEGe) and oxygen ($O_2$) were used as the Ge and O precursors, with flow rates of 160 sccm ($1.35 \times 10^{-5}$ mol/min) and 2000 sccm ($8.94 \times 10^{-2}$ mol/min), respectively. Argon (Ar) served as both the carrier and shroud gas at a flow rate of 1500 sccm. A detailed summary of the growth parameters is provided in Table 1.

Table1. Growth parameters for $GeO_2$ films deposited on sapphire substrates by MOCVD.

| Growth time (min) | Temp. (°C) | Pressure (Torr) | Rotation (RPM) | TEGe (sccm) | Shroud gas (sccm) | Oxygen (sccm) | Ar (sccm) |
|---|---|---|---|---|---|---|---|
| 10 | 925 | 300 | 300 | 160 | 2800 | 2000 | 1500 |
| 30 | 925 | 300 | 300 | 160 | 2800 | 2000 | 1500 |
| 60 | 925 | 300 | 300 | 160 | 2800 | 2000 | 1500 |
| 90 | 925 | 300 | 300 | 160 | 2800 | 2000 | 1500 |
| 10 | 880 | 300 | 300 | 160 | 2800 | 2000 | 1500 |
| 30 | 880 | 300 | 300 | 160 | 2800 | 2000 | 1500 |
| 60 | 880 | 300 | 300 | 160 | 2800 | 2000 | 1500 |
| 90 | 880 | 300 | 300 | 160 | 2800 | 2000 | 1500 |

The structural properties of the films were analyzed through high-resolution X-ray diffraction (HR-XRD) using a Bruker D8 DISCOVER system, which is equipped with a Cu K$\alpha_1$ radiation source ($\lambda = 1.5406$ Å), a triple-axis channel-cut monochromator, and an Eiger R 250K area detector. To investigate the surface topography and morphology, a Bruker Dimension ICON atomic force microscope (AFM) was used alongside a Quanta 600F environmental scanning electron microscope (SEM), which was integrated with micro-scale energy-dispersive X-ray spectroscopy (EDS) capabilities.

**Results and analysis**

Figure 1 illustrates the different polymorphs of $GeO_2$: amorphous, α-quartz, and rutile phases. Figure 1(a) shows the disordered structure of glassy $GeO_2$, where tetrahedral $GeO_4$ units are interconnected by oxygen atoms, forming a network without long-range periodicity. The glass phase is the kinetically easiest form of $GeO_2$, readily formed during thermal oxidation of elemental Ge. This formation occurs due to the high valence state (+4) of germanium, which bonds with oxygen atoms to create an interconnected network that is energetically close to the rutile phase.

As with SiO$_2$ and BO$_3$, GeO$_2$ is a strong glass former. In contrast, the crystalline phases of GeO$_2$ are represented in Figs. 1(b) and (c). Figure 1(b) shows α-GeO$_2$ (quartz), exhibiting trigonal symmetry with tetrahedral GeO$_4$ units arranged similarly to quartz, belonging to space group p3$_1$2$_1$. Figure 1(c) depicts rutile-type GeO$_2$, where GeO$_6$ octahedra are arranged in a tetragonal symmetry, forming a more compact structure, as seen in the space group p4$_2$/mnm. The commercially available source of GeO$_2$ is the quartz phase, which plays a key role in enabling the synthesis of GeO$_2$ films with diverse properties and applications.

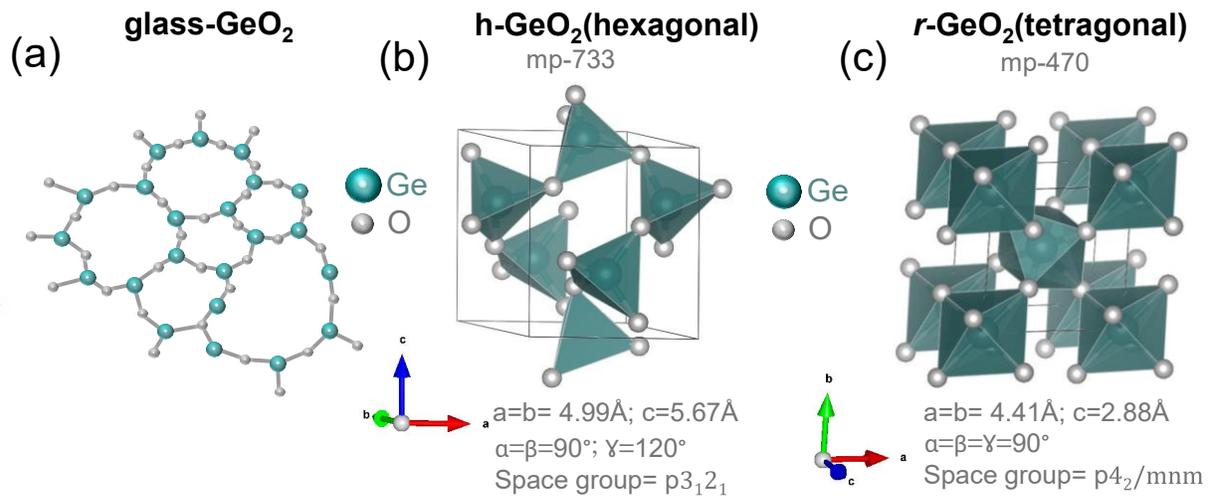

**Figure 1.** Crystal structures of GeO$_2$ polymorphs: (a) Amorphous GeO$_2$, with tetrahedral GeO$_4$ units forming a disordered network. (b) α-GeO$_2$ (quartz-like), space group p3$_1$2$_1$ with tetrahedral GeO$_4$ units arranged in a trigonal structure. (c) Rutile-type GeO$_2$, space group p4$_2$/mnm (mp-470), with octahedral GeO$_6$ units in a tetragonal arrangement.

Figure 2 describes the XRD $2\theta$ scans of GeO$_2$ thin films grown on r-plane [Fig. 2(a)] and c-plane [Fig. 2(b)] sapphire substrates at different growth durations (30, 60, and 90 minutes). For the r-plane samples [Fig. 2(a)], a clear progression is observed in the intensity of the diffraction peaks, which increases with the increment of growth time. Distinct reflections corresponding to the (100), (102), (110), (200), (211), and (203) planes of the hexagonal (quartz) GeO$_2$ phase become more prominent in the 60- and 90-minute samples. Despite the increasing sharpness of the peaks over time, the overall intensities remain relatively low while the full width at half maximum (FWHM) is broad, indicating a high density of structural defects. These defects are likely linked to the observed surface cracking. In the case of c-plane sapphire substrates [Fig. 2(b)], diffraction peaks corresponding to (100), (110), (211), and (203) planes of the hexagonal GeO$_2$ phase become increasingly distinct with longer growth times. Some diffraction signals are partially obscured by peaks from the substrate, but the general trend indicates crystallization behavior similar to that of the r-plane samples. The comparison of the films grown on both sapphire orientations reveals that, despite differences in surface morphology and defect density, GeO$_2$ films on r-plane and c-plane sapphire substrates adopt a similar polycrystalline quartz-phase structure and follow comparable growth behavior.

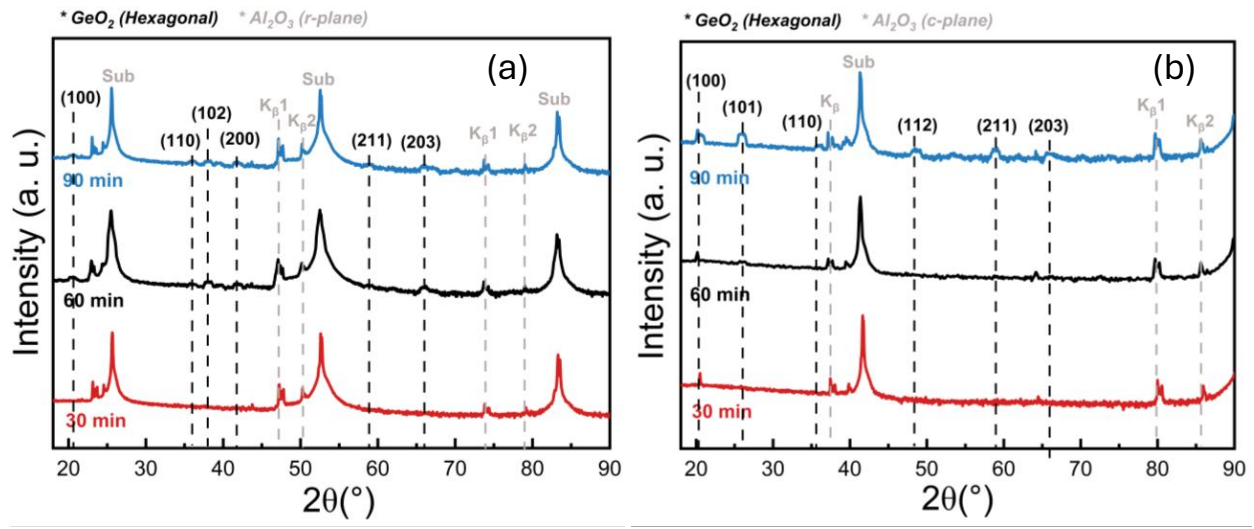

**Figure 2.** XRD *2θ* scans of GeO$_2$ thin films grown on (a) r-plane sapphire and (b) on c-plane sapphire.

The SEM images in Figure 3 illustrate the evolution of GeO$_2$ crystal growth on r-Al$_2$O$_3$ and c-Al$_2$O$_3$ substrates at two different temperatures: 925°C [Fig. 3(a)-(h)] and 880°C [Fig. 3(i)-(p)], with growth durations of 10 min, 30 min, 60 min, and 90 min, which reveal how temperature and growth duration influence crystal morphology. At 925°C, higher thermal energy promotes faster crystal growth, resulting in larger structures with intricate fractal-like patterns, including cracks that radiate outward. In contrast, at 880°C, the crystals are smaller, showing more compact growth patterns. The elevated temperature at 925°C or 880 °C enables enhanced coordination between Ge$^{4+}$ and O$^{2-}$ atoms, facilitating the formation of the crystalline quartz phase of GeO$_2$. During the initial stages of growth (10 and 30 minutes), the surface remains mostly amorphous, with partial crystallization occurring in certain regions. As the growth time increases, the area covered by spherulitic or leaf-like patterns expands. At 925°C, the size of these patterns increases from ~20 μm at 10 minutes to approximately 120 μm at 30 minutes, and further to ~400 μm at 60 minutes. After 90 minutes, the spherulitic patterns reach sizes larger than 550 μm and start to merge, forming a continuous pattern. This growth behavior is consistent on both r-$Al_2O_3$ and c-$Al_2O_3$ substrates, indicating that the increase in spherulite size with growth duration is independent of the substrate orientation. At 880°C, similar growth behavior is observed, but with smaller crystals, suggesting that lower temperatures result in slower crystal growth. Our previous studies have indicated that a significant crystallization driving force is required to form spherulitic patterns from the liquid phase, overcoming the energy barriers associated with nucleation and crystal growth [15]. The spherulitic patterns, which grow radially from a central core, exhibit non-regular patterns, often showing cracks in the center. These cracks likely result from the stress generated at the film-substrate interface, driven by the mismatch in thermal expansion coefficients between $GeO_2$ and $Al_2O_3$. This mismatch, combined with stress from negative volume changes (density of quartz > density of amorphous) during crystallization [24], contributes to the formation of cracks that radiate outward from the core. During crystallization, the stress from volume changes induces dislocation rearrangements, leading to lattice rotations, a phenomenon commonly observed in similar amorphous-to-crystalline transformations [22,25]. This rotating lattice behavior, often linked to dislocation dynamics and stress-induced volume changes during crystallization, further explains the observed growth patterns [26].

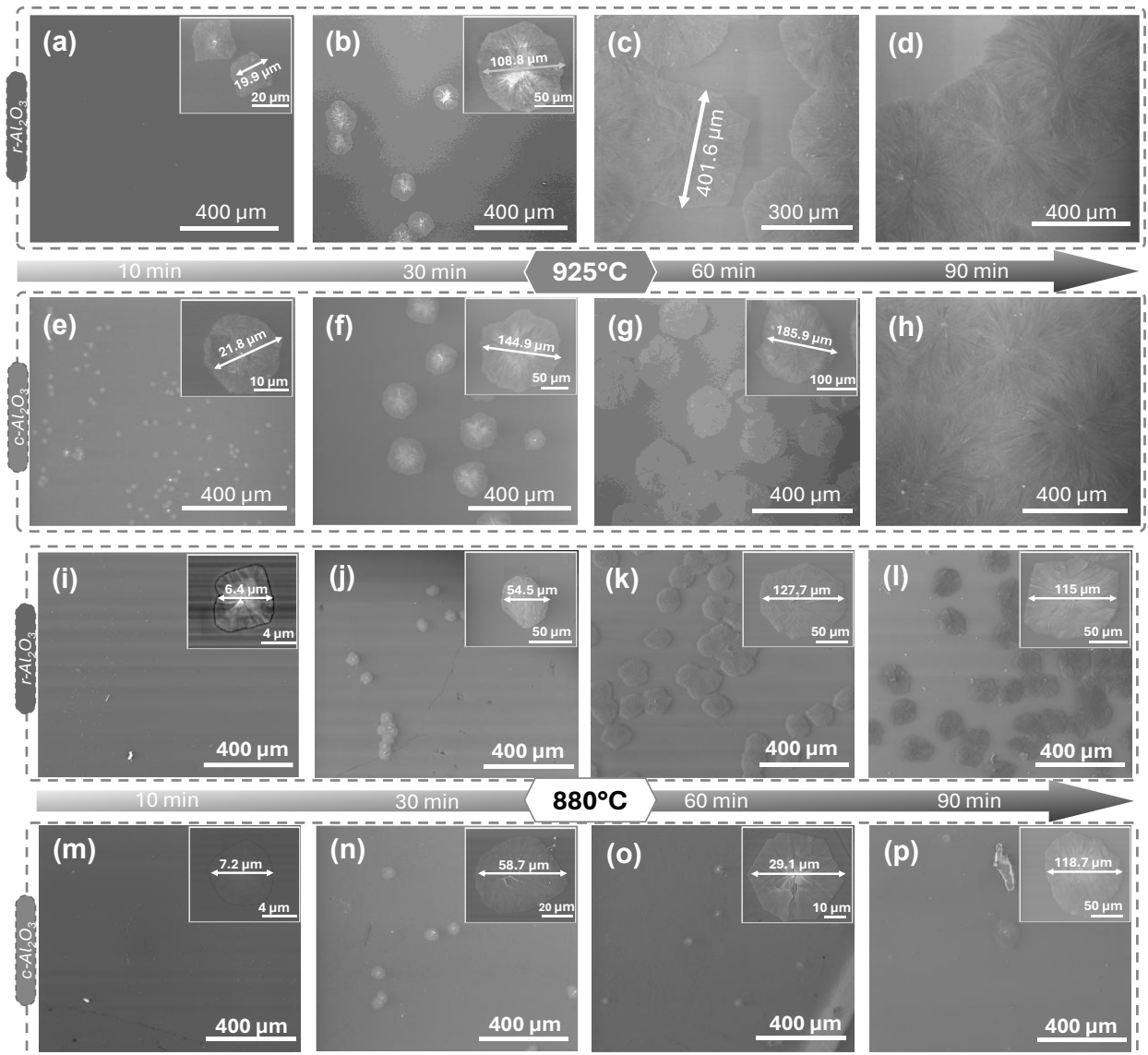

**Figure 3.** SEM images of GeO$_2$ crystal growth on sapphire substrates: (a–d) r-Al$_2$O$_3$ and (e–h) c-Al$_2$O$_3$ substrates at 925°C at different growth durations: 10 min, 30 min, 60 min, and 90 min. Insets show higher magnification images highlighting the lateral size of individual crystals. (i–l) r-Al$_2$O$_3$ and (m–p) c-Al$_2$O$_3$ substrates at 880°C at different growth durations: 10 min, 30 min, 60 min, and 90 min. Insets show higher magnification images highlighting the lateral size of individual crystals.

Even at the earliest stages of growth (10 minutes), the spherulitic patterns begin to emerge, indicating that the thickness of the film does not significantly impact their formation. In thin films, the interface energy becomes more significant than the volume energy, as growth is a two-dimensional (2D) process. This scaling of interface and volume energy terms, both proportional to

$r^2$ (the characteristic size of the structure), means that interface energy largely governs the morphology of the thin film, leading to the characteristic spherulitic patterns. Contamination particles can serve as nucleation cores for spherulitic patterns, facilitating the crystallization [26]. Our observed spherulitic patterns are generally irregular in shape, and their density is higher at the edges of the sample compared to the center. This suggests that nucleation is more favored at the edges, potentially due to defects or contamination particles introduced during the substrate cutting process. According to the dislocation model [27], dislocations generated at the glass/crystal interface form a cloud of geometrically necessary dislocations (GNDs), which are responsible for the bending of the crystal lattice during growth. This behavior may also be linked to the edges of the sample, where a thinner boundary layer could promote more pronounced lattice rotation and growth [15]. During crystallization, significant densification occurs along the glass/crystal interface, especially perpendicular to the free surface, where shape changes are less constrained [28]. This densification contributes to the formation of the observed spherulitic patterns. The irregular shape of the crack lines and fiber-like structures inside the spherulitic patterns highlights the anisotropic growth behavior at elevated temperatures, with crystal shapes reflecting the symmetry of quartz ($P3_121$ or $P3_221$)[29], as observed in the SEM images.

Figure 4 provides a quantitative analysis of the $GeO_2$ pattern growth, examining the evolution of the total crystal area [Fig. 4(a)], average crystal diameter [Fig. 4(b)], and surface coverage percentage [Fig. 4(c)] based on SEM images. The analysis incorporates a broad range of areas to ensure comprehensive calculations, and ImageJ software was used for the calculation. These measurements were conducted from Fig. 3 at different growth durations. In Fig. 4(a), the total crystal area increases significantly with time, especially at 925°C. As seen in Fig. 3, the $GeO_2$ films grown at this temperature exhibit rapid growth and the formation of distinct crystal patterns. The r-plane substrate shows the most pronounced expansion, reaching over 2.0 mm$^2$ at 90 minutes. In contrast, at 880°C, the area growth is suppressed, mirroring the more constrained pattern development observed at lower temperatures in Fig. 3. Figure 4(b) shows the evolution of the average crystal diameter. As in Fig. 3, the $GeO_2$ films at 925°C show rapid growth in crystal diameter, particularly within the first 60 minutes, with the r-plane substrate consistently yielding larger diameters than the c-plane. At 880°C, the diameter growth slows significantly, with the diameter plateauing after 30–60 minutes, indicating a slower crystallization rate at lower temperatures. In Fig. 4(c), the surface coverage evolves similarly, with a steep increase in coverage at 925°C, especially on the r-plane sapphire. However, the maximum coverage remains below 60% for all growth durations, indicating that longer growth times or post-growth annealing may be necessary to achieve full coverage.

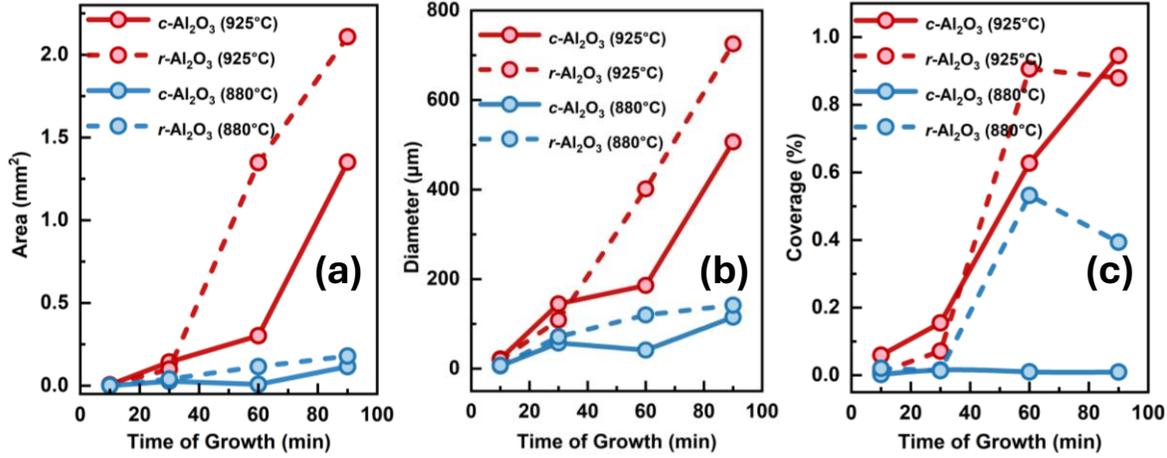

**Figure 4.** Growth duration evolution of (a) total crystal area, (b) average crystal diameter, and (c) surface coverage percentage for $GeO_2$ grown on c-plane $Al_2O_3$ and r-plane $Al_2O_3$ substrates at 925°C and 880°C.

Figure 5 demonstrates the influence of substrate symmetry on the morphology of $GeO_2$ crystals, as observed through structural schematics and SEM images. Figure 5(a) presents the r-$Al_2O_3$ surface, which exhibits a lattice configuration of a quadrangular pattern, while Fig. 5(b) shows the c-$Al_2O_3$ surface, characterized by hexagonal lattice pattern. These underlying substrate symmetries significantly affect the geometric characteristics of the $GeO_2$ crystals grown on these surfaces. In Fig. 5(c), the SEM image of a $GeO_2$ crystal grown on r-$Al_2O_3$ reveals a quadrangular-like morphology with ~90° internal angles, indicative of the lower in-plane symmetry of the r-plane sapphire. The $GeO_2$ crystal grows in an anisotropic manner, influenced by the distorted atomic arrangement of the r-plane substrate. Consequently, some $GeO_2$ patterns formed on r-plane sapphire are more elongated and irregular, reflecting the influence of the substrate symmetry on the crystal growth. Figure 5(d) shows the SEM image of a $GeO_2$ crystal grown on c-$Al_2O_3$, which adopts a hexagonal morphology with ~120° internal angles. This symmetry is characteristic of the hexagonal symmetry of the c-plane sapphire substrate. The sixfold symmetry of the c-plane substrate promotes more symmetric lateral expansion of the $GeO_2$ patterns, resulting in the formation of well-defined hexagonal domains.

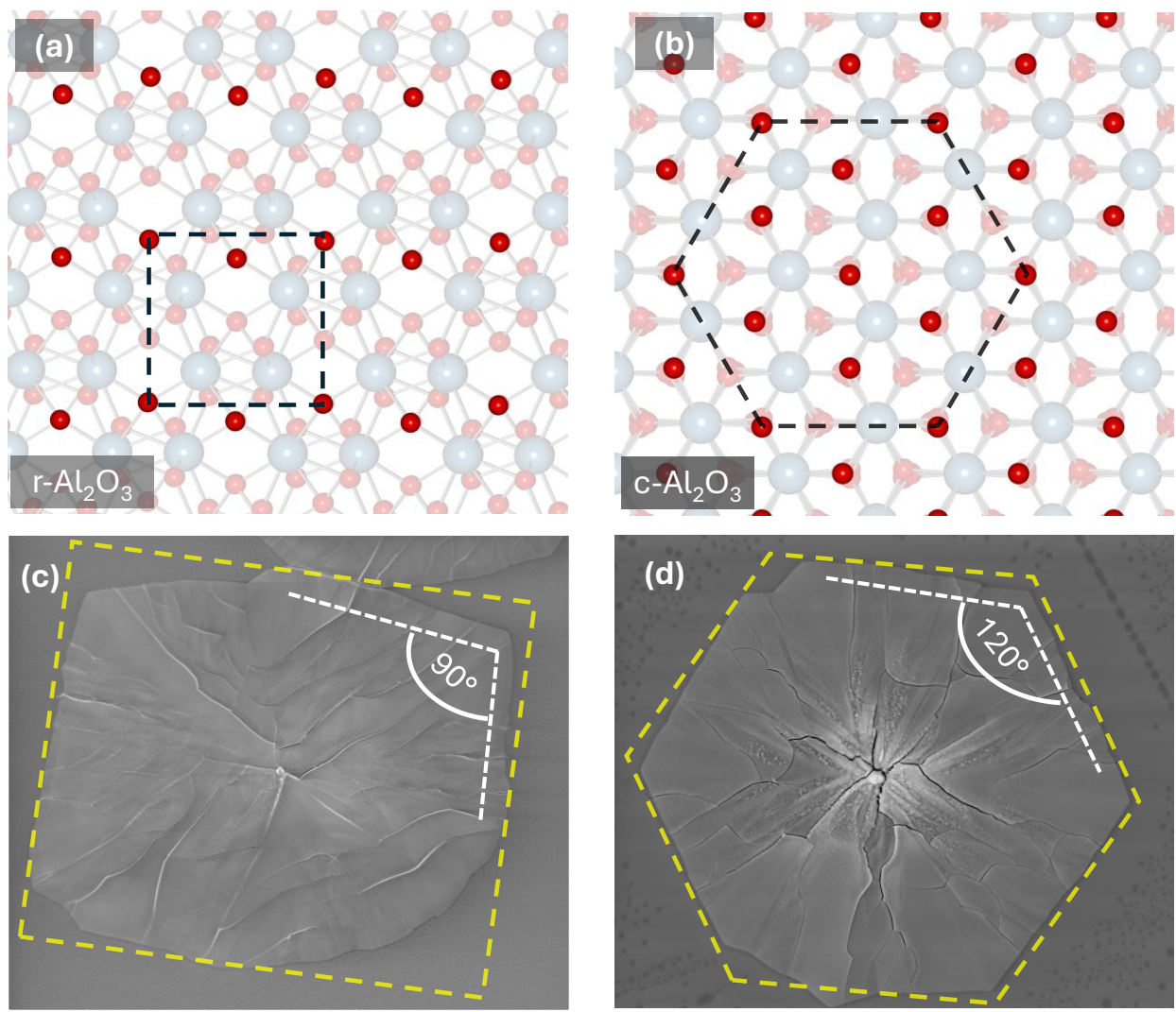

**Figure 5.** Structural schematics and SEM images illustrating the effect of substrate symmetry on GeO$_2$ crystal formation. (a) r-Al$_2$O$_3$ surface exhibiting square-like local symmetry and (b) c-Al$_2$O$_3$ surface exhibiting hexagonal symmetry. (c) SEM image of a GeO$_2$ crystal on r-Al$_2$O$_3$ showing quadrilateral-like morphology with 90° internal angles. (d) SEM image of a GeO$_2$ crystal on c-Al$_2$O$_3$ exhibiting hexagonal morphology with 120° internal angles.

Additionally, as observed in our study, the morphology of GeO$_2$ crystals can reflect the phase style. Compared with the patterns of quartz phase, the rutile phase of GeO$_2$ typically exhibits a square shape, while amorphous GeO$_2$ displays smooth regions without any distinct patterns [14,30]. This morphology difference can help us distinguish between different phases of GeO$_2$. In addition to the morphology by SEM, AFM measurements were employed to provide additional insights into the surface morphology. Figure 6(a-c) shows the evolution of surface morphology of the GeO$_2$ film on c-Al$_2$O$_3$ substrates at 10, 30, and 60 minutes of growth by AFM. As the growth period increases, morphology evolves from small, randomly distributed spherulitic structures to larger, more developed domains with noticeable changes in surface features. The increased lateral growth corresponds to a refinement of the surface, indicating the progression from an initial disordered state to more defined structures. A terrace structure, step-flow growth, can be seen on the pyramid-like GeO$_2$ island, especially shown in Fig. 6(b). Figure 6(e) presents line profiles taken across selected regions, showing an increase in both size and height as the growth duration extends. The 3D surface morphology in Fig. 6(d), derived from Fig. 6(c), further highlights the characteristic pyramid-like structure of the GeO$_2$ island and the step-flow growth on the island, indicating a mixture of 2D and 3D growth [19-21].

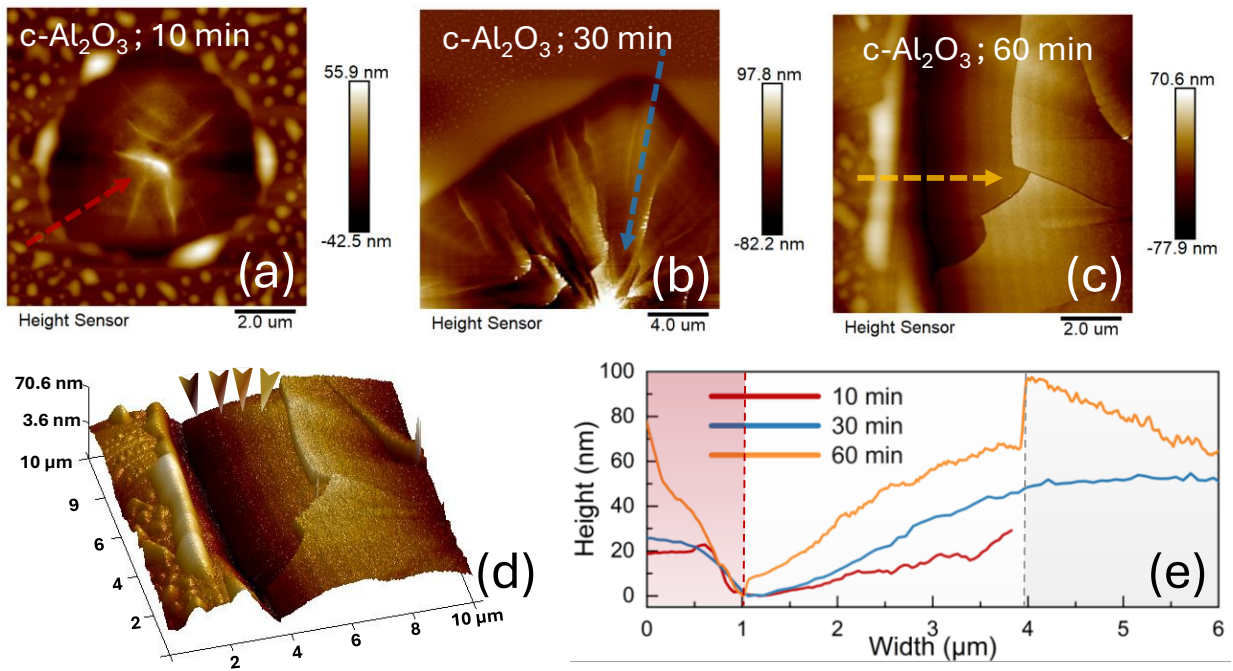

**Figure 6.** Atomic force microscopy (AFM) analysis of GeO$_2$ film evolution on c-Al$_2$O$_3$ substrates: (a–c) AFM images for growth times of 10, 30, and 60 minutes, showing progressive lateral growth and morphological changes. (d) 3D AFM image of (c) showing the formation of stair-like structures. (e) Line profiles across selected crystals reveal increasing terrace width and step height with extended growth duration.

The formation and evolution of pyramid-like patterns can be attributed to periodic accumulation and relaxation of epitaxial strain during lateral expansion, which leads to vertical height adjustments as the domains grow outward. Step boundaries are evident, offering insight into the transition from amorphous to crystalline $GeO_2$. The development of well-defined crystalline domains is contrasted with the more disordered, amorphous regions, revealing the progress of crystallization as the growth time increases. Such dividing can also be observed in the corresponding height profile of Fig. 6(e), which illustrates the distinct phase transitions between the amorphous and crystalline regions. The profile clearly shows the relative heights of the two phases with a valley, indicating the negative volume change during transition from the amorphous phase to the quartz phase.

Based on the findings above, Fig. 7 presents schematics illustrating the evolution of $GeO_2$ patterns on sapphire substrates are shown in Fig. 7, highlighting the combined effects of growth temperature and growth duration on the pattern formation. The top side of the figure demonstrates the temperature effect, starting from an amorphous film that gradually transforms into crystalline $GeO_2$ patterns as the temperature increases, while neither film nor patterns are found for temperatures below 725 °C. For the same growth period (10 min), higher temperature promotes the crystallization of the quartz phase. On the bottom side, the growth duration effect is depicted. At shorter growth durations, the patterns remain small and less defined. As the growth time increases, the quartz structures begin to emerge and the patterns grow larger, indicating the lateral growth and enhanced crystallization of $GeO_2$. A pyramid-like morphology also develops during this growth period, with the central region continuing to grow vertically while the edges expand laterally in a layer-by-layer manner, resulting in a stepped structure along the sidewalls. This schematic illustrates how temperature and duration jointly influence the morphology of $GeO_2$ films, with higher temperatures promoting faster crystallization and longer durations leading to the development of larger, more intricate patterns.

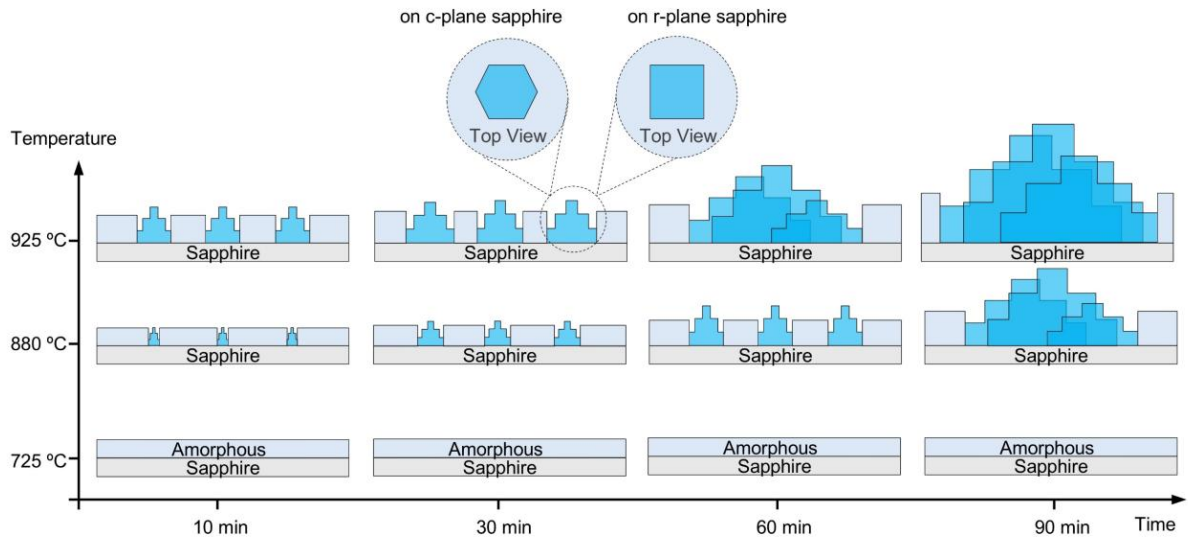

**Figure 7.** Schematics of the effects of growth temperature (also based on results from our previous study [15]) and growth duration on evolution of $GeO_2$ grown on sapphires.

**Conclusion:** This study presents the growth behavior and morphological evolution of $GeO_2$ films on sapphire substrates, focusing on the effects of growth temperature and duration on the phase transition and nucleation of the quartz phase $GeO_2$. XRD analysis confirms the progressive crystallization of the quartz phase with increasing growth time on both c-plane and r-plane sapphire substrates. SEM imaging reveals the temporal evolution of characteristic surface patterns, demonstrating that elevated temperatures significantly accelerate the quartz phase formation. AFM measurements further resolve the morphology of pyramid-like $GeO_2$ islands, exhibiting step-flow features during the phase transition. These features also suggest a volume contraction associated with the amorphous-to-quartz transformation. Notably, no rutile-phase $GeO_2$ was detected under the studied MOCVD conditions, suggesting a narrow growth window for the rutile phase on sapphire for MOCVD. These findings can help advance the understanding of phase evolution, nucleation mechanisms, and growth dynamics of $GeO_2$ thin films synthesized by MOCVD.

## AUTHOR DECLARATIONS

### Conflict of Interest

The authors have no conflicts to disclose.

### Author Contributions

Botong Li: Data curation (lead); Investigation (equal); Methodology (lead); Formal analysis (supporting); Writing – review & editing (supporting). Imteaz Rahaman: Formal analysis (lead); Investigation (equal); Writing-original draft (lead). Hunter D. Ellis: Writing – review & editing (supporting); Investigation (supporting). Bobby Duersch: Data curation (supporting). Kathy Anderson: Methodology (supporting). Kai-Fu: Conceptualization (lead); Writing – review & editing (lead); Supervision (lead); Project administration (lead); Resources (lead).

**ACKNOWLEDGEMENT**

The authors would like to express their sincere gratitude for the financial support received from the University of Utah's start-up fund. Furthermore, they acknowledge the instrumental facilities provided by the University of Utah, which include the Utah Nanofab Cleanroom, the Material Characterization Meldrum, and the Nanofab Electron Microscopy and Surface Analysis Facilities.

**DATA AVAILABILITY**

The data that supports the findings of this study are available from the corresponding authors upon reasonable request.